\documentclass[jpcm,showpacs,twocolumn,showemail]{revtex4}
\usepackage{bbm}
\usepackage{mathrsfs}
\usepackage{epsfig,psfrag}
\usepackage{amsmath,amsfonts,amssymb}
\usepackage[usenames]{color}
\usepackage[dvips, bookmarks, colorlinks=true, plainpages = false, citecolor = blue, linkcolor = blue, urlcolor = blue, filecolor = blue]{hyperref}
\def\be{\begin{equation}}
\def\ee{\end{equation}}
\def\bea{\begin{eqnarray}}
\def\eea{\end{eqnarray}}

\begin{document}

\preprint{draft}

\title{Linear Relationship Statistics in Diffusion Limited Aggregation}
\author{ Abbas Ali Saberi}\email{a$_$saberi@ipm.ir}

\address {School of Physics, Institute for Research in Fundamental
Sciences (IPM), P.O.Box 19395-5531, Tehran, Iran}
\date{\today}

\begin{abstract}
We show that various surface parameters in two-dimensional diffusion
limited aggregation (DLA) grow linearly with the number of
particles. We find the ratio of the average length of the perimeter
and the accessible perimeter of a DLA cluster together with its
external perimeters to the cluster size, and define a microscopic
schematic procedure for attachment of an incident new particle to
the cluster. We measure the fractal dimension of the red sites
(i.e., the sites upon cutting each of them splits the cluster) equal
to that of the DLA cluster. It is also shown that the average number
of the dead sites and the average number of the red sites have
linear relationships with the cluster size.
\end{abstract}


\maketitle

Diffusion-limited aggregation (DLA) is a model of a growing cluster,
originally proposed by Witten and Sanders \cite{witten}. The model
has been shown to underlie many pattern forming processes including
dielectric breakdown \cite{db}, electrochemical deposition
\cite{ed}, viscous fingering and Laplacian flow  \cite{vf}
\emph{etc}. It is defined by a simple stochastic model on a square
lattice as follows. A \emph{seed} particle is located at the center
of the lattice, and then a random walker is released from infinity
$-$ operationally, from a point of radius much larger than the
radius of the growing cluster. Upon contacting, the random walker
sticks irreversibly to the cluster. Repeating the process, leads to
an intricate ramified structure whose surface in the plane grows
proportional to the bulk (this will be shown in this letter).\\This
procedure is equivalent to solving Laplace's equation outside the
aggregated cluster with appropriate boundary conditions. In two
dimensions, since analytic functions automatically obey Laplace's
equation, the theory of conformal mappings provides another
mechanism for producing the shapes. This method has been directly
used by Hastings and Levitov to study DLA \cite{HL}.\\One of the
most interesting aspects of such an aggregate is the multifractal
behavior of the growth site probability distribution (the harmonic
measure) $\{p_i\}$, where $p_i$ is the probability that the site
$i$, belonging to the perimeter of the cluster, will grow at the
next time \cite{HMP,HDH}. The screened sites with tiny growth
probability, play an important role in determining the
multifractality, while evaluating the harmonic measure on these
sites is too difficult. The theoretical difficulty emerges from
solving equations with boundary conditions on a complicated growing
interface.

Different numerical methods applied for large aggregates in plane,
show that DLA does not follow a simple fractal pattern and deviates
from linear self-similarity \cite{MKA}. However most of the studies
have been focused on the scaling behavior of different quantities in
DLA, here in this letter, we show that there exist couple of linear
relationships which, to our knowledge, have not been addressed yet.
We show that the perimeter and the accessible perimeter of a growing
aggregate, together with their externals grow linearly with the
cluster size $n$, with different rates. The number of sites which
are not accessible for the incident random walker will be shown to
increase linearly with size. Our results indicate that the almost
particle attachments (which a particle is assumed to be at the
center of a square of the mesh size $a$), choose \emph{single-side}
contact. We also use the concept of \emph{red sites}, as a measure
of ramification of the aggregates and compute their fractal
dimension. The average number of the red sites is another quantity
which has a linear relationship with the number of aggregated
particles.

\begin{figure}[!b]\begin{center}
\includegraphics[scale=0.8]{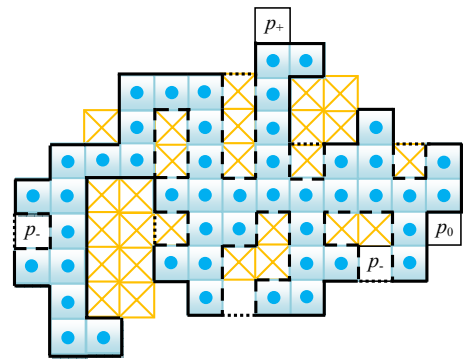}
\narrowtext\caption{\label{Fig1}(Color online) A putative small DLA
cluster of $n$ aggregated particles shown in shaded squares. The
marked squares with $\times$ show the dead sites which are
inaccessible for the random walker coming from infinity. The hull of
the DLA, defined in the text, is the union of the solid and dashed
lines. The outer perimeter is the union of the solid and dotted
lines. The accessible perimeter is the hull of the union of the DLA
cluster and the dead sites. The ($n+1$)-th incident random walker
may have either positive, negative or zero contribution to the
length of the hull, denoted by $p_+$, $p_-$ or $p_0$, respectively.
}\end{center}
\end{figure}

In order to investigate the behavior of the aforementioned
quantities, we simulated several independent on-lattice DLAs of
different mass up to $n=10^5$ particles. One can see that an
ensemble of simulated DLAs is strongly fluctuating, and to obtain
consistent results the averages have to be taken over a large number
of samples. In this letter, the averages are taken over $5000$
independent samples for different cluster sizes.

The first quantity that we measure is the perimeter or \emph{hull}
of an aggregate. Consider a cluster of aggregated particles on a
square lattice (see Fig. \ref{Fig1}), where each frozen particle in
the cluster is assumed to be located at the center of a plaquette.
To define the hull, a walker moves clockwise around the aggregate
and along the edges of the corresponding lattice starting from a
given boundary edge of the cluster. The direction at each step is
always chosen such that walking on the selected edge leaves a frozen
particle at the \emph{right} and an empty plaquette at the
\emph{left} of the walker. If there are two possible ways, the
preferable direction is one that is on the right of the walker. The
directions \emph{right} and \emph{left} are defined locally
according to the orientation of the walker. This algorithm yields
the hull of the DLA whose length $l$ (in the units of the mesh size
$a$), is equal to the number of steps until the walker returns to
the starting edge.

\begin{figure}[h]\begin{center}
\includegraphics[scale=0.45]{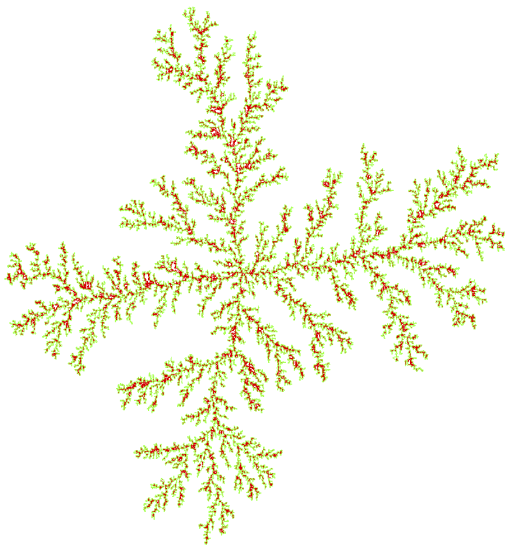}
\narrowtext\caption{\label{Fig2}(Color online) The hull (dark
colored) and the accessible perimeter (light colored) of a DLA
cluster of size $n = 10^5$. }\end{center}
\end{figure}

\begin{figure}[!b]\begin{center}
\includegraphics[scale=0.4]{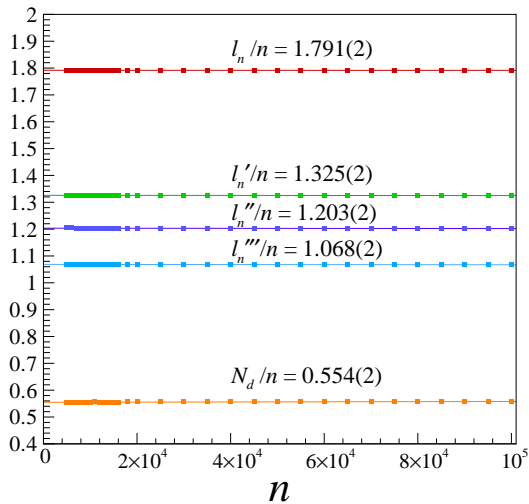}
\narrowtext\caption{\label{Fig3}(Color online) From top to bottom:
the rescaled length of the perimeter $l_n$, external perimeter
$l^{'}_n$, accessible perimeter $l^{''}_n$, accessible external
perimeter $l^{'''}_n$, and rescaled number of the dead sites $N_d$
versus the cluster size $n$. The errors are less than the symbol
size. The solid lines indicate the best linear fit of zero
slope.}\end{center}
\end{figure}

Let us denote the length of the hull of a cluster of size $n$, by
$l_n$. We now show that $l_n$ must be proportional to the cluster
size. \\To have a cluster of size $n$+1, a random walker is released
from infinity. It sticks to an edge of the hull with the probability
proportional to the harmonic measure there. We define the three
possibilities as shown in Fig. \ref{Fig1}, during which the random
walker can stick to the cluster having either \emph{single},
\emph{double} or \emph{triple}-\emph{side} contacts. Upon selecting
each of them, it may have positive, negative or zero contribution to
the length of the updated hull. According to its contribution, we
denote the probability that the random walker selects each of the
three possibilities as $p_+$, $p_-$ or $p_0$, respectively. It can
be easily checked that the events with single side contacts have
always a positive contribution of $+2$ to the length of the hull,
while the two other possible events can either have zero or negative
contribution of $2n_-$, with $n_-\geq 1$. One can indeed define the
length of the updated hull by considering these contributions to the
previous length $l_n$, using the following recursive relation
\be\label{Eq1} l_{n+1}=l_{n}+2 [p_+-n_- p_-].\ee As our following
experiments suggest, for $n\gg 1$, the last term in the above
relation seems to be independent of the cluster size and therefore,
one can obtain that $l_n\sim 2 [p_+-n_- p_-]n$. Our simulation
result for the average length of the hull $l_n$ as a function of the
cluster size $n$ is in a good agreement with this linear
relationship, see Fig. \ref{Fig3}. We find that
$\frac{l_n}{n}=1.791(2)$. This yields the infimum of the probability
that the incident random walker chooses a \emph{single side}
contact, and hence we obtain $p_+ \gtrsim 90\%$.

In addition to studying the hull of the aggregates, we also study
the external perimeter of the hull which can measure the number of
sites got trapped in the fjords (proportional to $n_-$). To define
the external perimeter, we first close off all the narrow
passageways of a lattice spacing on the DLA cluster and then looking
at the hull of the resulting cluster (see Fig. \ref{Fig1}). We find
experimentally that the length of the external perimeter $l^{'}_n$
has also a linear relationship with the cluster size $n$. The best
fit to our data shows that $\frac{l^{'}_n}{n}=1.325(2)$.

All sites on the hitherto considered perimeters are not necessarily
accessible by the incident random walker in all regions of the
aggregate. Therefore, it is of interest to measure the totally
accessible perimeter which is, in principle, a \emph{hull} that
surrounds the union of the DLA cluster and all inaccessible sites by
the incident random walker coming from infinity$-$ see Fig. \ref{Fig1}.\\
In order to determine the accessible perimeter, we introduce an
algorithm which seems to be more efficient than one used in ref.
\cite{AMS}. This may be called '\emph{burning algorithm}', during
which each accessible site around the cluster will be marked as a
\emph{burning} site, and all cluster sites and inaccessible ones
will be left unmarked.

The algorithm begins by drawing a box that includes the entire
cluster without touching it, and 'marking in' from the boundary of
the box. A boundary site of the box is selected and is marked as
'\emph{free}' if there is not any cluster site at its nearest
neighborhood and as '\emph{burnt}' otherwise. If the site left
'free'-marked, after checking all of its nearest neighbor sites and
marking each of them as 'free' or 'burnt' as before, it is recolored
as 'burnt'. Repeating this procedure for all 'free'-marked sites and
all unmarked boundary sites of the box, partitions inside the box
into two regions, one containing the sites which are being marked as
'burnt' sites and are all accessible by the random walker, and the
other region i.e., inaccessible region, is the union of the cluster
sites and unmarked sites (or \emph{dead} sites) which are not
accessible by the random walker. The accessible perimeter is then
the hull of the inaccessible region.\\
We find that the average length of the accessible perimeter
$l^{''}_n$, and the average number of dead sites $N_d$, grow
linearly with the cluster size $n$. The linear relations can be
obtained as shown in Fig. \ref{Fig3}, according to
$\frac{l^{''}_n}{n}=1.203(2)$ and $\frac{N_d}{n}=0.554(2)$,
respectively.

Following the same ratiocination as Eq. (\ref{Eq1}) for these
observations, one can estimate the probability that the incident
particle sticks to the accessible perimeter of DLA with a
single-contact. We denote the same probabilities as before that the
incident random walker coming from infinity to stick to the
accessible perimeter with \emph{single}, \emph{double} or
\emph{triple}-\emph{side} contacts by $p'_+$, $p'_-$ or $p'_0$,
respectively. Note that for the hull of accessible perimeter, all
\emph{single}, \emph{double} or \emph{triple}-\emph{side} contacts
always have positive, zero or negative contribution (of $+2$, $0$ or
$-2$, i.e., $n_{-} = 1$) to the length, respectively. Unlike DLA
perimeter (because of screening with the dead sites), all sites on
the accessible perimeter have nonzero growth probability. It can be
again shown that $l''_n\sim 2 [p'_+ - p'_-]n$. This result is
checked in Fig. \ref{Fig3}, according which we obtain that the
length of the accessible perimeter $l''_n$ grows linearly with the
cluster size $n$ as $\frac{l''_n}{n} = 1.203(2)$. It can thus be
inferred that the incident square-like random walker attaches to the
active zone of the cluster with probability of $p'_+ \gtrsim 60\%$
by choosing the single-side contact. Since all fjords of narrow
throat are filled by the dead sites inside the accessible perimeter,
so we can estimate the number of sites $\mathcal{N}_-$ on it with
triple-side contact possibility, by closing off all the narrow
passageways of a lattice spacing. This yields the outer of the
accessible perimeter whose length $l'''_n$ has also linear
relationship with the size of he cluster, i.e., as shown in Fig.
\ref{Fig3}, $\frac{l'''_n}{n} = 1.068(2)$. Comparing with the same
relation for $l''_n$, one can conclude that $\mathcal{N}_- \sim
0.067 n$. However, by closing off the narrow passageways, all
possible remaining contacts will be single or double-side ones,
nevertheless it is not possible to estimate an exact value for
$p_+$. It is because that these attachments can also have negative
contribution to the length of the external accessible perimeter.

\begin{figure}[t]\begin{center}
\includegraphics[scale=0.4]{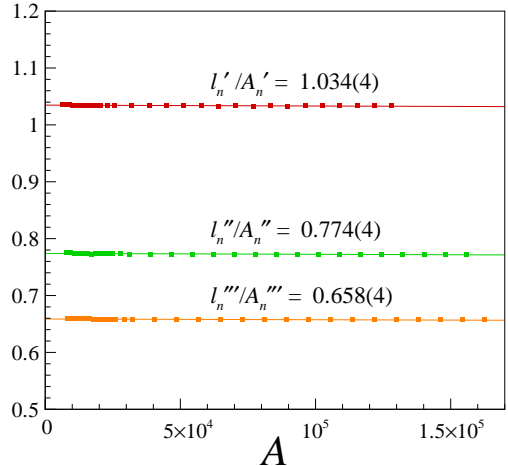}
\narrowtext\caption{\label{Fig4}(Color online) Linear relationships
between different perimeter lengths and the area enclosed by them.
From top to bottom: the length of the external perimeter $l^{'}_n$,
accessible perimeter $l^{''}_n$ and accessible external perimeter
$l^{'''}_n$, rescaled by their enclosed area denoted by $A'_n$,
$A^{''}_n$ and $A^{'''}_n$, respectively. }\end{center}
\end{figure}

These linear relationships seem to be the characteristic features of
DLA clusters. In fact, for common fractals appear in two-dimensional
statistical mechanics, such as critical Ising or percolation
clusters, the length of the cluster boundaries (or loops) $l$ with
the fractal dimension of $d_f$, has a scaling relation with the area
of the loops as $A\sim l^{2/d_f}$.\\Motivated by this relation, we
examine the behavior of the length of the perimeters versus the area
enclosed by them. As depicted in Fig. \ref{Fig4}, we find that all
the perimeter lengths $l^{'}_n$, $l^{''}_n$ and $l^{'''}_n$, have
linear relations with their area. Note that this area is not
necessarily equal to the cluster size (multiplied by the square
lattice spacing $a^2$).\\These linear relationships show that
applying the scaling relation $A\sim l^{2/d_f}$ to measure the
fractal dimension of the perimeter of DLA clusters would thus lead
to a misleading result reported in \cite{L}.

The fractal dimension of the perimeters $d_f$ can be computed by
using the scaling relation between the average length of the
perimeter $l$ and a linear size scale, e.g., the gyration radius
$R_g$, i.e., $l\sim R_g^{d_f}$. As long as the gyration radius of
the growing cluster is used as the linear size scale in this
relation, we find that the fractal dimension of all perimeter
lengths, within the statistical errors, are equal with the same
value as the fractal dimension of DLA cluster. This result is in
agreement with the same one reported in \cite{AMS}. Nevertheless, if
we scale the perimeter lengths with their gyration radius (i.e., the
gyration radius of the loops produced by the perimeters themselves),
we find a minor difference among the fractal dimensions. If $d^c_f$,
$d_f$, $d^{'}_f$, $d^{''}_f$ and $d^{'''}_f$ represent the fractal
dimension of the DLA cluster, the perimeter, the external perimeter,
the accessible perimeter and the accessible external perimeter,
respectively, we find that $d^c_f = 1.707(3)$, $d_f = 1.710(3)$,
$d^{'}_f = 1.717(3)$, $d^{''}_f = 1.723(3)$ and $d^{'''}_f =
1.725(3)$.

\begin{figure}[t]\begin{center}
\includegraphics[scale=0.4]{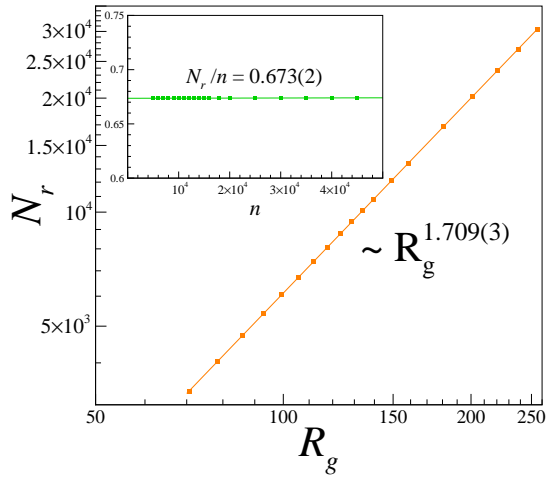}
\narrowtext\caption{\label{Fig5}(Color online) Main frame: the
average number of red sites $N_r$ versus the gyration radius $R_g$
of DLA clusters of different sizes. Inset: linear relation between
$N_r$ and the cluster size $n$.}\end{center}
\end{figure}

In the following, we use the concept of the \emph{red sites}
borrowed from two-dimensional critical structures \cite{HS}, to have
a quantitative measure for the ramification of DLA cluster. A red
site on the DLA cluster denotes a site that upon cutting leads to a
splitting of the cluster. This measures the number of nodes
connected by effectively one-dimensional links.

We carried out simulations in order to compute the fractal dimension
of the red sites $d_r$ on DLA cluster by using the scaling relation
$N_r \sim R_g^{d_r}$, where $N_r$ is the number of red sites and
$R_g$ is the gyration radius of the DLA cluster. Due to the large
amount of time, the simulation was performed for clusters of size
$n\leq 4.5\times 10^4$. As shown in Fig. \ref{Fig5}, we find that,
within statistical errors, the fractal dimension of the red sites is
equal to that of the DLA cluster, i.e., $d_r = 1.709(3)$.\\ We also
find that the number of the red sites grows linearly with the
cluster size according to the relation $\frac{N_r}{n} = 0.673(2)$.

In conclusion, we found that different surface parameters on DLA
cluster such as the length of the hull, the accessible perimeter and
their externals and also the number of dead sites grow linearly with
the cluster size. These have been used to investigate the
microscopic feature of the cluster growth by measuring the
probability that a square-like incident random walker attaches to
the cluster by choosing either the single, double or triple-side
contacts. We also found that the border of DLA grows linearly with
the total area enclosed by it. We have measured the fractal
dimension of the red sites on the DLA cluster and found it to be
equal with that of the cluster itself. The average number of the red
sites has been shown to have a linear relationship with the cluster
size.

\end{document}